\documentclass[fleqn,11pt]{article}

\def\philprivate#1{{\color{magenta}[[#1]]}} \def\philprivate#1{} 
\let\oldmarginpar\marginpar \renewcommand\marginpar[1]{\-\oldmarginpar{\raggedright\footnotesize\color{green} #1}}
\def\includefigs{y}

\newcommand\CUT[1]{}
\usepackage{afterpage}
\usepackage{scisupplement} 
\usepackage{epsfig}
\usepackage{graphicx}
\usepackage{pdfsync}
\usepackage{outlines}
\usepackage{mathpazo}
\usepackage{fullpage}
\def\pnlabel#1{{\color{DkRed}\bfseries\small\label{#1}}}
\def\figstatus#1{\philprivate{#1}}
\def\Nspeedfigure{$10\,\msunit$} 
\def\Nnoisedef{$-30\,\uVunit$} 
\def\Nspikedef{$-40\,\uVunit$} 
\def\NinterestingVThresh{$-44\,\uVunit$} 

\def\NfitCorLen{0.18\,\msunit}
\def\NfitCorMag{57\,\uVunit^2}\let\Neta=\NfitCorMag
\def\Nx{0.58} 
\def\Nfitspikes{$1\,260\,475$}

\def\NIFpercent{0.02}

\def\Nduration{120\thinspace min}

\def\mh{\mu} 
\def\Nfall{\xi}  
\def\Ncov{\mathsf{C}} 

\def\bV{\mathbf{V}}   
\def\bF{\mathbf{F}}   
\def\xbar{\bar x}\def\ybar{\bar y}  

\def\NinterestingVThreshsym{V^*_{\rm trust}}

\def\sref#1{Sect.~\ref{s:#1}}    
\def\fref#1{Fig.~\ref{f:#1}} 
\def\eref#1{Eqn.~\ref{e:#1}}

\def\Nitem#1{\par
\hangindent=1.3em\noindent\hbox to 1.3em{\textsl{#1}}}
\def\Nfirstitem#1{\par\hangindent=1.3em\noindent\hbox to 1.3em{#1.}}

\def\draftfig{{\sc draft}}
\def\optics{{\sc optics}}
\def\<#1>{\texttt{#1}}
\newcommand{\ex}[1]{{\mathrm e}^{#1}}                 

\def\bivector#1{\raise1.5ex\hbox{\vphantom{m}$\leftrightarrow$}\mkern-16.5mu #1\,}
\usepackage[usenames]{color}\definecolor{DkRed}{cmyk}{0,.5,.5,.3}
\def\dd{\mathrm{d}}

\newcommand{\inv}{^{\raise.15ex\hbox{${\scriptscriptstyle-}$}\kern-.05em 1}}
\def\ml{\textsl{Matlab}}

\def\msunit{\ensuremath{\mathrm{ms}}}

\def\mmunit{\ensuremath{\mathrm{mm}}}
\def\umunit{\ensuremath{\mu\mathrm{m}}}

\def\kHzunit{\ensuremath{\mathrm{kHz}}}
 
\def\uVunit{\ensuremath{\mu\mathrm{V}}}

\newcommand{\capitem}[1]{(\textit{\uppercase{#1}})~}
\def\capcolor#1{\textit{#1}}
\def\yesflag{y}

\title{Fast, scalable, Bayesian spike identification for multi-electrode arrays}
\date{}
\author{Jason S.\ Prentice${}^1$, Jan Homann${}^1$, Kristina D.\ Simmons${}^2$, Ga\v{s}per Tka\v{c}ik${}^1$,\\
Vijay Balasubramanian${}^{1,2}$, and Philip C. Nelson${}^1$\\ {}\\
${}^1$ Department of Physics and Astronomy, ${}^2$ Department of Neuroscience, \\University of Pennsylvania, Philadelphia, PA 19104, USA
}

\begin{document}
\maketitle
\begin{abstract}
	We present an algorithm  to identify individual neural spikes observed on high-density multi-electrode arrays (MEAs). Our method can distinguish large numbers of distinct neural units, even when spikes overlap, and accounts for intrinsic variability of spikes from each unit.  As MEAs grow larger, it is important to find spike-identification methods that are \emph{scalable}, that is, the computational cost of spike fitting should scale well with the number of units observed. Our algorithm accomplishes this goal, and is fast, because it exploits the spatial locality of each unit and the basic biophysics of extracellular signal propagation.  Human intervention is minimized and streamlined via a graphical interface.  We illustrate our method on data from a mammalian retina preparation and document its performance on simulated data consisting of spikes added to experimentally measured background noise.   The algorithm is highly accurate.
\end{abstract}

\section*{Author summary}
Single neurons transmit messages in the form of electrical pulses called ``spikes.'' Networked populations of neurons in the brain can use patterns of spikes to encode information or perform computations.  To decipher this neural code, we must record simultaneously from many cells.  The primary tools for such measurements are grids of closely spaced electrodes called multi-electrode arrays.  The challenge in using these probes is that signals from single neurons typically appear on several electrodes, while each electrode records signals from multiple neurons.  Disentangling these signals to determine which neurons fired, and when, is a principal bottleneck in understanding the collective behavior of neural circuits.  Here we present an efficient and accurate Bayesian approach to this ``spike sorting'' problem that can scale to arrays of hundreds of electrodes.   Our techniques accommodate variability in spike waveforms and identify spikes correctly even when nearby neuronal responses obscure one another.  The key is  judicious use of theory: we systematically model and exploit the spatial characteristics of signal propagation and the structure of noise within a simple Bayesian model of neural activity.

\section{Introduction\label{s:intr}}
The vertebrate retina is an important model system in neuroscience because it is amenable to detailed study despite having a complex structural and functional architecture \cite{Peterreview}.    Population coding and collective behavior in the retinal output is studied by use of multi-electrode arrays (MEAs) to record extracellularly from many retinal ganglion cells (RGCs) simultaneously \cite{Meister:1994p1395,Devries:1997p2762}.  Similar recordings can now also be made in other brain areas \cite{Buzsaki:2004p1417}.  MEAs offer unprecedented possibilities to obtain both single neuron and single action potential resolution from large tissue samples. However, recordings obtained in this way are useful only if every spike can be correctly assigned to the neuron that generated it. Even if each neuron spikes with a unique waveform signature, we must still determine all those ``template'' waveforms present in a dataset, separating them from each other and from noise. Moreover, in practice there can be wide variation in the spike waveforms from a given unit (for instance in amplitude), complicating the task of determining from data which units fired and when. 

Solving this ``spike sorting problem'' is the principal bottleneck in the use of high density arrays with hundreds or thousands of electrodes.  Methods that were manageable with tetrodes \cite{Gray:1995p2550} do not generally scale up to the massive datasets that large arrays generate. For example, some standard methods cluster data by manually examining two-dimensional projections in a feature space of a few tens of dimensions. This approach is infeasible when the feature space contains thousands of dimensions.

Another challenge with large arrays is that the chance of seeing a single isolated spike becomes negligible, simply because there is so much activity. Thus we must find template waveforms corresponding to the activity of single neural units without ever seeing a pristine example of one, and we must be prepared to decompose temporally overlapping spikes in essentially every recorded event. Overlaps in both space and time are less frequent, but they nevertheless must be resolved if we wish to unravel the patterns of collective neural activity.   Resolution methods that rely on exhaustively checking all possible combinations suffer a combinatorial explosion for large arrays.  Further, any spike decomposition method must stop before every spike has been found, because there will be some units whose intrinsic amplitude is not larger than recording noise.  We need a principled approach to terminating each fit and to deciding later which units' activities have been reliably captured.

Thus, to be useful for large arrays, a spike identification algorithm must both \emph{scale well} and also \emph{be able to decompose overlapping events.}  This article outlines a method that accomplishes these goals. It first clusters a small subset of a larger dataset, using an automated ordering technique combined with rapid human cluster-cutting.   This manual step is efficient, and scalable, because (i) the ordering arranges event data by similarity along a single dimension, (ii) the ordered data display band-like features that are visually very salient for human operators, making cluster cutting unambiguous,  and (iii) the algorithm is robust to variations and outliers in the cluster-cutting procedure. The algorithm then fits the full dataset to the spike templates thus obtained, using a modified Bayesian approach.   In our data (from guinea pig retina) most of the intrinsic variability of spikes from a given unit consists of amplitude variation only, whereas other variability can be summarized as a universal (spike-independent) noise process.  By carefully modeling these circumstances we greatly reduce our computational burden. 

After characterizing the spatiotemporal character of the noise, our algorithm identifies spikes iteratively in a matching-pursuit (or ``greedy'') scheme \cite{Mallat:1993p2651,Segev:2004p1133}. Fitting terminates when addition of another spike does not improve the likelihood score of a fit; no \emph{ad hoc} complexity penalty is needed.   No assumptions are made about spike firing times or  cross-correlations; in particular, we do not require \emph{a priori} any refractory ``hole'' in the spike time autocorrelation functions. Nevertheless, all of the inferred spike trains corresponding to otherwise acceptable spike types do exhibit such a hole, which serves as a check on our results.  Any automated clustering algorithm requires human proofreading; we structure our methods (and develop tools) to make this as easy as possible.

\begin{figure}[tb!]
\begin{minipage}[b]{\textwidth}
\ifx\includefigs\yesflag
\begin{center}
\hspace{0in}
\includegraphics[width=6.83 in]{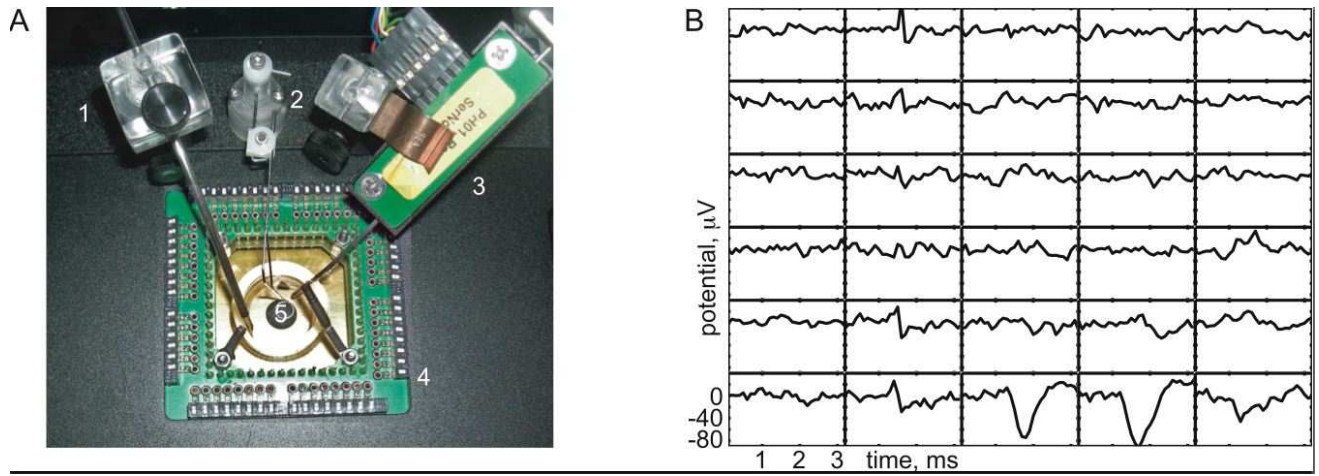}
\end{center}\fi
\end{minipage}
\hfil\renewcommand{\baselinestretch}{.95}\normalsize \smallskip \caption{\small
 \capitem{a}{}Typical MEA apparatus. A tissue sample was mounted in an inverted microscope, with images projected onto it via a small video monitor at the camera port (not visible). \capcolor{Clockwise from left,} 1: suction; 2: tissue hold-down ring; 3: perfusion inflow, with temperature control; 4: preamplifier; 5: location of the multi-electrode array. 
 \capitem{b}{} Example of a single-spike event. Each subpanel shows the time course of electrical potential ($\uVunit$) on a particular electrode in the $5\times6$ array. The electrodes are separated by $30\,\umunit$ (similar to RGC spacing). A spike from one unit is visible in the lower right corner and an axonal spike can be seen running vertically in the second column of electrodes. Data were acquired at $ 10\,\kHzunit$. After baseline subtraction and high-pass filtering (see supplement), a spatial whitening filter was applied (see \sref{pp}). \pnlabel{f:MEA}}\par
\end{figure}%

Our approach combines successful elements from previous techniques:  the empirical characterization of the noise \cite{Pouzat:2002p1281}; separation of clustering and fitting steps and the iterative subtraction scheme for handling overlaps \cite{Segev:2004p1133}; and division of the clustering task by leader electrode address \cite{Litke:2004p1134}.  Novel features of our approach include  systematic exploitation of the spatial organization of the signals, the use of an ordering algorithm to greatly simplify clustering, the observation that the noise temporal correlation is well represented by a simple function, the characterization and use of a prior distribution on spike amplitude variation, and the introduction of a principled  Bayesian likelihood criterion for terminating spike fitting.   Each of these innovations adds a critical element to the success of our spike sorting method.  Although we focused on data taken on vertebrate retina, the methods should be equally applicable to other kinds of MEA data, for example in other brain areas \cite{Buzsaki:2004p1417}.

\section{Results\label{s:res}}

To illustrate our method, we tested our spike sorting algorithm on \Nduration{} of recordings from guinea pig retinal ganglion cells (RGC), acquired with a 30-electrode, dense MEA covering about $0.018\,\mmunit^2$ of tissue (\fref{MEA}A). The analysis described in this paper identified \Nfitspikes\ spikes in the dataset.  A typical firing event took the form \fref{MEA}B, where each panel shows 3ms of the electrical potential recorded by each electrode (or ``channel'').  We identified spiking events as voltages crossing a threshold of \Nspikedef, taking into account the fact that simultaneous threshold crossings on neighboring channels represent the same spike event  (see Methods for details). The duration of each spike event was taken to be $3.2$ ms centered on the event's peak. 

In addition to identifiable spikes, each electrode had background activity with a standard deviation of  $\sim$10 $\mu V$  that we will collectively refer to as ``noise.''  Potential sources for this activity include true (Johnson) noise in the electrode and electronics, electrical pickup from the environment, as well as a hash of background activity from weak or distant neurons \cite{Fee:1996p1393}. A challenge for spike identification is that in general there is no way to separate these three classes of ``noise'' cleanly  from each other, nor from the spikes of interest to us.   Nevertheless, we will propose a technique for identifying spikes that is very accurate for firing events with intrinsic amplitude at least 4 times  the standard deviation of the noise.  
\fref{MEA}B  illustrates that each single firing unit will be ``heard'' on multiple electrodes, and that those electrodes form a spatially localized group. Our method is scalable because it systematically exploits this simple observation: even on a large electrode array, most firing units will involve only a handful of electrodes.  (Some of our signals were {\it not} local, and stretched over the entire electrode array in a line (e.g., \fref{MEA}B).  We ignored such axonal firing events, which were also distinguished by their low amplitude and triphasic shape.)

\subsection{Preliminary visualization of our data\label{s:gpd}}
We first attempted a ``geographical clustering'': from each event we found the minimum of the potential on each channel and the channel containing the deepest minimum (``leader channel'').  We then used the absolute values of the minima as weights in a weighted average of the locations of the $9$ electrodes neighboring the leader channel. This weighted average gave a particularly salient two-dimensional feature, the event's {\it barycenter} $(\xbar,\ybar)$. Taking the third feature $z$ to be minus the absolute peak potential gave a scatterplot that clearly showed many well-separated clusters (\fref{3plot}A), without any attempt to deduce the best selection of three features by principal component analysis (\fref{3plot}B).    This extension of the ``triangulation'' method developed for tetrode recordings \cite{Gray:1995p2550} already shows key aspects of the data: (a)~many clusters are highly dispersed in amplitude, and (b)~some cluster pairs appear at nearly the same spatial locations but are nevertheless well separated by amplitude. The first circumstance means that we must allow for variable amplitude when fitting spikes to templates representing the clusters. The second warns us that a simple least-squares fit to amplitude could confound two distinct units. For this reason our spike-fitting method creates a Bayesian prior for each cluster's amplitude variation, allowing us to make such discriminations.

\begin{figure}[tb!]
\begin{minipage}[b]{\textwidth}
\ifx\includefigs\yesflag
\begin{center}
\hspace{0in}
\includegraphics[width=6.83in]{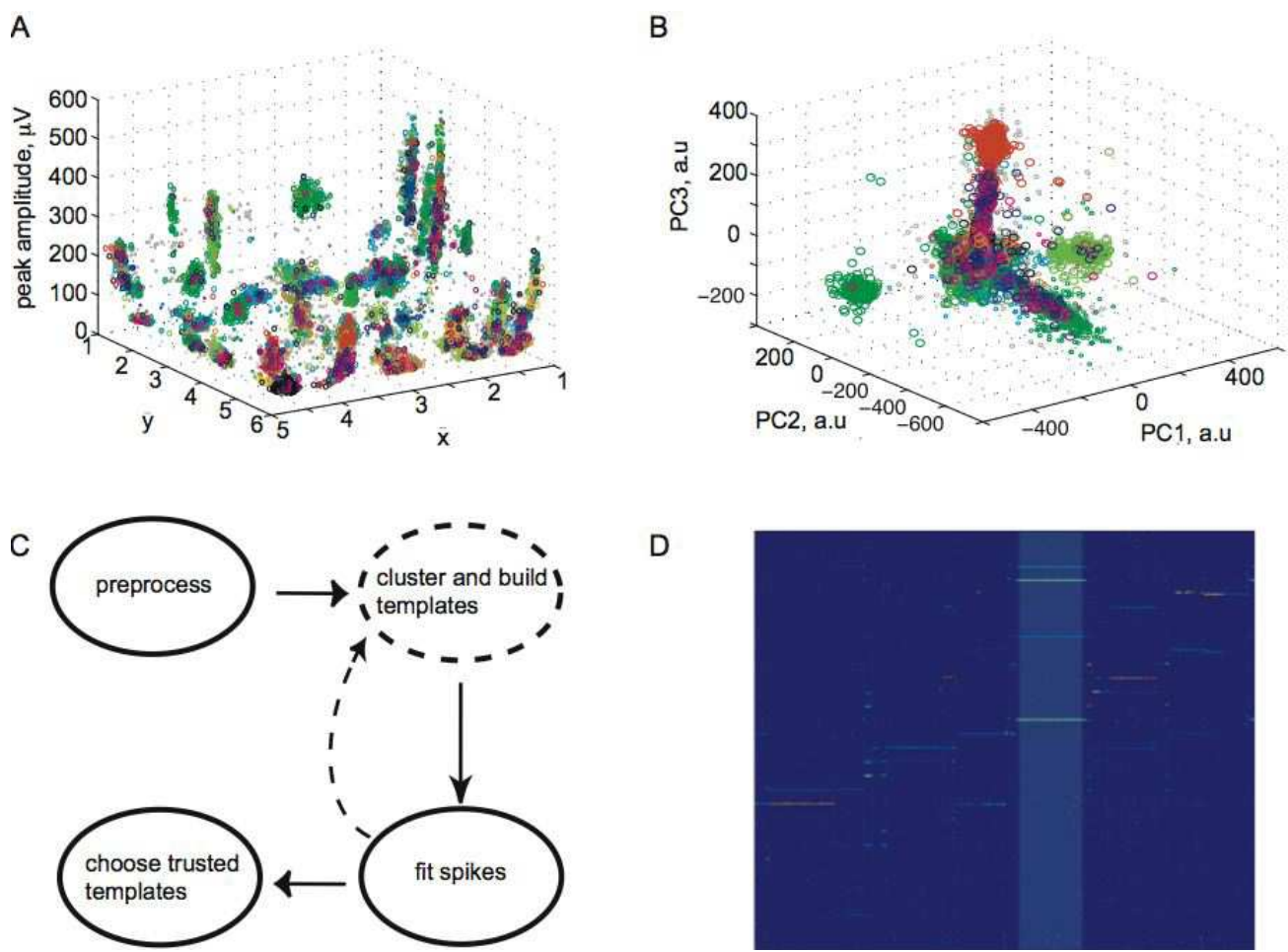}
\end{center}\fi
\end{minipage}
\hfil\renewcommand{\baselinestretch}{.95}\normalsize \smallskip \caption{\small
 \capitem{A}{}  $22\,234$ firing events cluster well in terms of their {\it barycenter} (voltage weighted average spatial location) and absolute peak voltage (see text), despite wide amplitude dispersion in some groups; each combination of color and marker size corresponds to one spiking unit identified by the clustering method developed in the text. Grey points were unassigned to any cluster. A total of 107 clusters are marked. 
\capitem{B}{}  Events  cluster poorly when projected onto the three principal features uncovered by principal component analysis (PCA). Coloring as in (A).
\capitem{C}{} Schematic of our spike sorting method.   Dashed lines involve a small subset of a full dataset. The backwards arrow describes the introduction of new spike templates found after the first pass of fitting (\sref{sf}); a total of two passes are performed. 
\capitem{D}{} The \optics{} algorithm orders all firing events into a linear sequence based on a distance measure (see text).  Events are lined up in this order (x-axis), and represented in terms of the 960 voltage samples recorded by all the electrodes during a 3.2 ms firing event (y-axis; from top to bottom, 32 consecutive time samples from one channel, then 32 time samples from the second channel, and so on).  The human operator  highlights bands of events (typically very clear to an observer) that appear to constitute a single cluster; one such band is shown.  Later automated diagnostics refine and check these assessments.  \pnlabel{f:3plot}}\par
\end{figure}%

Although the simple clustering based on spatial location in \fref{3plot}A looks promising, it can be misleading. Indeed, the restriction of the weighted average to the $9$ electrodes around the leader can  artificially separate clusters by biasing the barycenter to be located near a particular electrode. This problem could be alleviated by using a larger neighborhood, but on large arrays there will inevitably be temporal collisions of spikes from distinct units.  The barycentric features in \fref{3plot}A will register such collisions as a haze of seemingly random spots.   Thus, at a minimum the MEA voltage traces must be segmented by exploiting the spatial locality of recorded responses.  Despite these shortcomings, \fref{3plot}A points out why other, more sophisticated, methods can succeed: the ``geographic'' information encoded by the MEA is a powerful intrinsic clue to each unit's identity.

\subsection{Summary of our method}
Our sorting method is outlined in \fref{3plot}C (details in Methods).    From a subset of the raw data, we made a preliminary classification of spike events in terms of the electrode on which they achieved their peak voltage. All events sharing a given leader channel were cropped to the $9$ electrodes neighboring the leader, then ordered with the \optics{} algorithm \cite{Aanke99b} into a linear sequence.  The \optics{} algorithm places similar events nearby in the sequence; distance was measured by  a normalized Euclidean distance between event voltage traces (see Methods).     The linear sequence of events was displayed to the user along with all the recorded voltage samples for each event (\fref{3plot}D), and manually clustered. Although the ordering was based on events cropped to $9$ channels, the full waveforms were displayed to the user (\fref{3plot}D). Because the data are ordered in one dimension, and because precision is not required in view of later automated refinement, this manual step remains rapid. An automated method for cluster cutting could be implemented, but in view of the inevitable need for human proofreading we preferred to simply carry out this step by hand.  From each preliminary cluster, we estimated a template waveform representing the corresponding neural unit and then fit the templates to the remaining data.   Fitting was accomplished by a Bayesian algorithm based on a probabilistic model capturing the dominant sources of variability we observed in our data: background noise, spike amplitude, and overlapping spikes from distinct units.  After finding, for each event, the most probable template which accounts for the event, we subtracted it and then iterated. Finally, the fit results were used in a post-hoc validation of the initial clustering, and we repeated the procedure in a second pass if necessary.  Details of each step are presented in Methods.

\subsection{Tests of our method\label{s:tests}}

\optics{}-based clustering of our dataset led to 107 potential templates for events from distinct neural units.     Many of these templates had low amplitudes; such low-amplitude templates were sometimes mistakenly fit to noise by our algorithm. We  therefore rejected units that were likely to contain substantial noise fits because they were of amplitude less than or comparable to the noise (details in \sref{ecr} and \fref{testnoise}D).  This left fifty potentially reliable units  that were accepted in our dataset.

\paragraph{Comparison with geographical clusters: }  
Our OPTICS-based procedure identified 107 potential clusters of events in a subset of the data.    To check that the procedure gave reasonable results we plotted each event in the barycentric coordinates of \fref{3plot}A, and colored the events according to the cluster label.   The clusters were spatially localized and separated in peak amplitude, as they should be if they were produced by distinct single neurons.    Gray dots in \fref{3plot}A were not assigned to any cluster.   Some of these events contained overlaps of spikes that were not resolved by the initial spatial segmentation of data during the preprocessing step.  The subsequent spike fitting step in our algorithm attempted to resolve such ambiguities.

\paragraph{Error rates on synthetic data: }  To validate our algorithm we tested its performance on synthetic data created by adding spikes to experimentally measured background noise clips, then fitting templates to each clip. We took noise clips to be $3.2$ ms segments of time during which no spikes were recorded on any channel; we identified 14\,000 such clips.   For each clip, we randomly chose a fixed number (1, 3, 5, 7, or 9) of templates from the initial set of 107, with uniform probability and without replacement. We then added these templates to the noise clip at random times, leaving a margin of $0.6$ ms on either side of the clip to prevent waveforms from being cut off. (Our typical template waveforms extended approximately $0.5$ ms to either side of the peak.)  We gave each spike an amplitude drawn from a Gaussian distribution with mean equal to its template amplitude and standard deviation 10\% of the template amplitude (this was similar to the observed distribution). The template fitting algorithm was then run over this synthetic dataset and analyzed for false positive and false negative rates (\fref{OffCellCenters}A,B). We counted a false negative for a template every time that template was present in an event but not fit correctly to within $1$ ms; we counted a false positive every time a template was fit to the data without actually being present.  The error rates increased with the number of template overlaps; thus, for the fifty templates with amplitudes that exceed the noise,  we separately plotted error rate histograms for each degree of overlap.   Error rates were robustly low --- even within extremely complex events with 9 overlapping spikes (very rare in the data), the majority of spike templates had an error rate of a few percent or less.  To gain perspective on these values, we measured the number of templates fit to each event in our recorded data: 60\% of events contained 1 spike, 25.5\% 2 spikes, 8.5\% 3 spikes, 3\% 4 spikes, 1\% 5 spikes, and 2\% had more than 5 spikes.    Most of the errors were made on lower amplitude templates for which amplitude variations can lead to confusion with noise.

\paragraph{Refractory violations: } 
 When sorting spikes recorded extracellularly, ground truth can be assessed if simultaneous intracellular recordings are available, e.g., \cite{Harris:2001p2555}.  Since we do not have such recordings, in order to validate our algorithm on real data we examined the rate of refractory violations --- i.e., the fraction of  interspike intervals of duration less that 1.5 ms.  Refractory violations can appear in our sorted data if spikes from distinct neural units are mis-assigned to same unit, or if noise fluctuations are mistaken for spikes.   Of the 107 templates constructed from the initial clustering 84\%  had less than 0.5\% refractory violations and all had less than 2.5\%, providing evidence that the templates produced by the initial clustering rarely merge distinct neural units (\fref{OffCellCenters}C).    All fifty templates describing units that rose reliably over the noise level had less than 0.5\% refractory violations. Futhermore,  $96\%$  of these  had less than 0.1\% refractory violations (\fref{OffCellCenters}C).   This is strong evidence that our algorithm makes few fitting mistakes on the units otherwise identified as reliable.

\begin{figure}[tb!]
\begin{minipage}[b]{\textwidth}
\ifx\includefigs\yesflag
\begin{center}
\hspace{0in}
\includegraphics[width=6.83in]{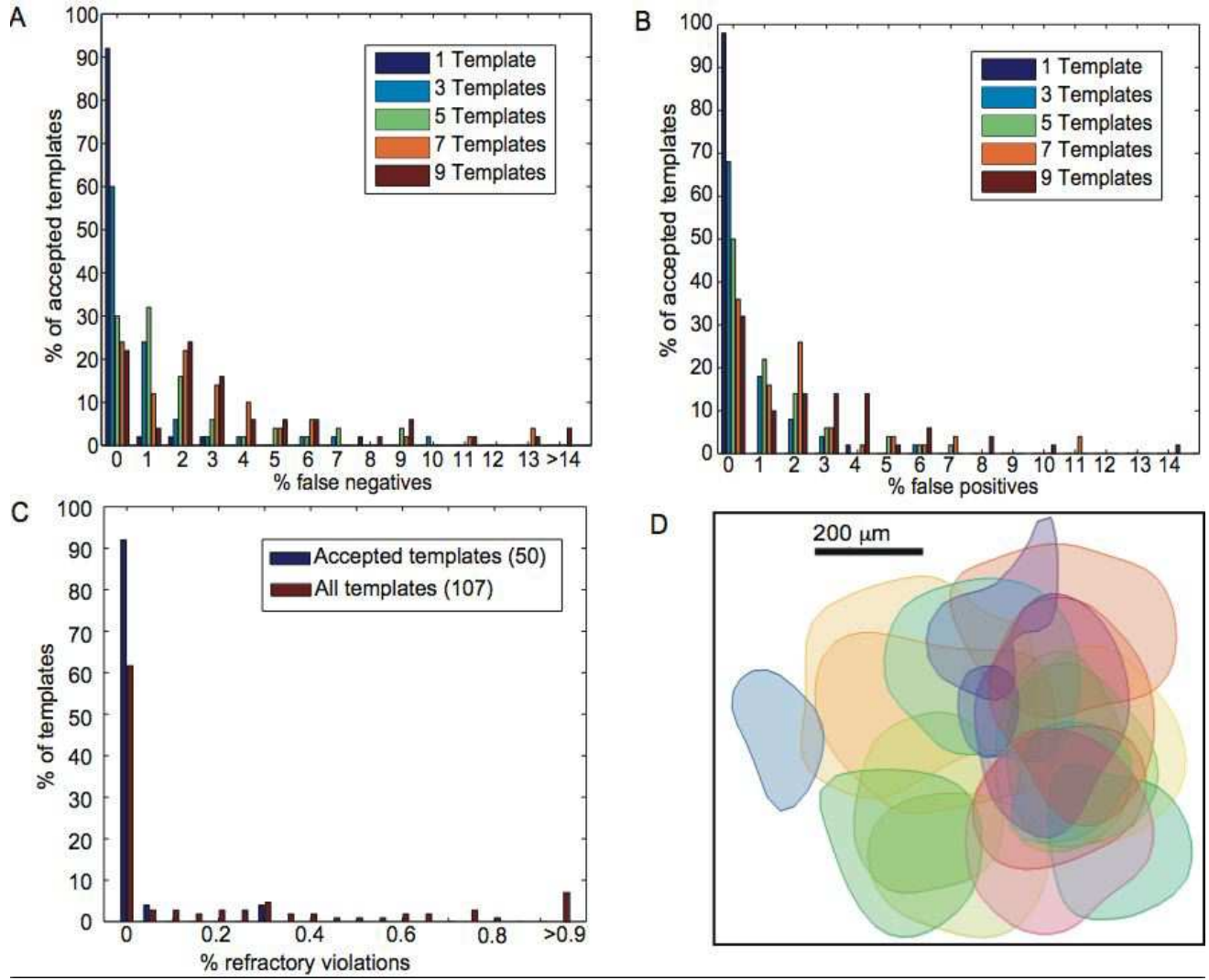}
\end{center}\fi
\end{minipage}
\hfil\renewcommand{\baselinestretch}{.95}\normalsize \smallskip \caption{\small
\capitem{a} The percentage of templates with different false negative fractions in fits to synthetic data (fraction of times a fit was {\it not} reported for a template when it was actually present). 
\capitem{b}The percentage of templates with different false positive fractions in fits to synthetic data (fraction of times a fit was reported for a template when it was \emph{not} actually present).  Results in (A) and (B) reported separately for events with different numbers of template overlaps (inset colors).
\capitem{c} In fitting real data, refractory violations are rare (see text).
\capitem{d} The centers of 19 OFF cell receptive fields recorded from a single piece of tissue. To map a neuron's receptive field center, we first find the peak (in space and time) of the spike-triggered average stimulus. Restricting to the peak time, we apply cubic spline interpolation in space and then draw contour lines at 75\% of the peak value.
 \pnlabel{f:OffCellCenters}}\par
\end{figure}%

\paragraph{Coverage: }  While the absence of refractory violations gives evidence that our algorithm does not merge different neural units together, it might still split spikes from the same unit into two distinct clusters if, e.g. there was substantial amplitude variation.  To test for this, for each unit that was above the noise level we measured the linear receptive field by taking the spike triggered average (STA) of the flickering checkerboard stimulus.    We expect that such receptive fields will be connected regions of the visual field, roughly elliptical in shape, and that no two units will have identical receptive fields.   31 of the 50 reliably identified units had enough spikes to give reliable estimates of the spatial receptive field.  Of these, examination of the temporal kernel showed that 19 were OFF cells (responding to light decrement) and 12 were ON cells (responding to light increment), consistent with the expected excess of OFF ganglion cells \cite{OnOff}.  None of these receptive fields were identical, giving evidence that our algorithm did not split single units into multiple clusters.   Further, all of the receptive fields were connected, suggesting that none of our clusters are mixtures of different RGC.  In addition, essentially all of the recorded area was covered by at least one receptive field (coverage of OFF cells shown in \fref{OffCellCenters}D).   The density of RGCs in guinea pig  varies from 250$\,\mmunit^{-2}$ to1500$\,\mmunit^{-2}$ \cite{DoNascimento:1991p2761}. We receive signals from a region slightly larger than the electrode array, roughly $0.065\,\mmunit^2$. Thus the expected number of RGC is 16--97, comparable to our total of 31 receptive fields, keeping in mind that many of the sluggish cell types would not have enough spikes to yield a good spike triggered average.

\begin{figure}[tb!]
\begin{minipage}[b]{\textwidth}
\ifx\includefigs\yesflag
\begin{center}
\hspace{0in}
\includegraphics[width=6.83in]{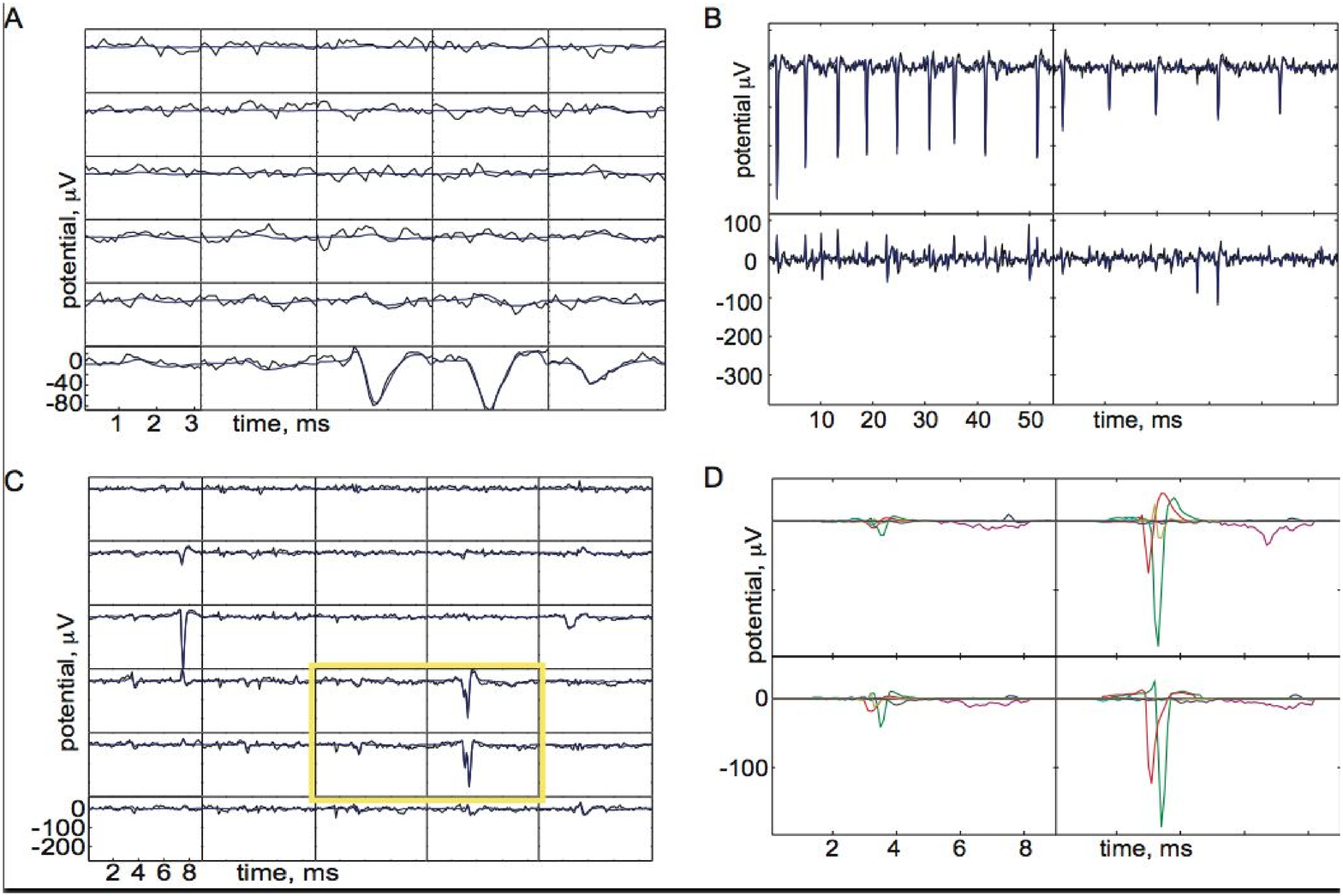}
\end{center}\fi
\end{minipage}
\hfil\renewcommand{\baselinestretch}{.95}\normalsize \smallskip \caption{\small
\capitem{A}{} Example of a single-spike event. Each subpanel shows the time course of electrical potential (in $\uVunit$, \capcolor{black curves}), on a particular electrode in the $5\times6$ array.   After baseline subtraction and high-pass filtering, a spatial whitening filter was applied (see Methods). \capcolor{Blue curves} show the result of our fitting algorithm, in this case a single template waveform representing an individual neural unit.
\capitem{B}{} Detail of a more complex event and its fit, in which a single unit fires a burst of 9 spikes of varying amplitudes (upper left channel), while a different unit fires 5 other spikes (upper right channel).
\capitem{C}{} Example of an overlap event and its fit, which now is a linear superposition of 7 templates.
\capitem{D}{} Detail of (C), showing signals on four of the electrodes. This time individual fit spikes are displayed. The \capcolor{red} and \capcolor{green} traces show fit templates that, although similar, differ significantly in their overall strength, and in the relative strengths of their features. The \capcolor{olive} trace shows a fit to a low-amplitude  template that was later classified as unusable, and hence was discarded, by the procedure in \sref{ecr}. 
 \pnlabel{f:expt}}\par
\end{figure}%

\paragraph{Complex events: } A major challenge for a spike sorting algorithm is dealing with variability in spikes produced by individual neural units.   An even greater challenge arises from spatio-temporal overlaps between spikes from different neural units.  Our low error rate in analysis of synthetic data containing both of these complexities (\fref{OffCellCenters}) provides evidence that our algorithm is effective at resolving overlaps and identifying variable spikes from given units.   To test this further, we manually examined many events in the real data which a human observer could identify as representing overlaps or neural variability; and the algorithm typically did an excellent job of dealing with  variable-amplitude bursts (\fref{expt}B), as well as events that overlap in space and time (\fref{expt}C).

\paragraph{Speed: } Currently the main fitting code, written in \ml, requires about \Nspeedfigure{} of real computer time per fit spike on a commercial 2.5\thinspace GHz computer, times 2 for the two passes. This is fast enough for our purposes; considerable further improvement is possible with existing software (Mex) and hardware (GPU) techniques.

\section{Materials and methods}

\paragraph{Ethics statement}

This study was carried out in accordance with recommendations from the National Institutes of Health and the guidelines of the American Veterinary Medical Association.  The protocol was approved by the Animal Care and Use Committee of the University of Pennsylvania (No. 803091).   All surgery was performed under ketamine/xylazine and pentobarbital anesthesia, and all efforts were made to minimize suffering.

\paragraph{Experimental procedure}
Our methods were developed and tested on retinal response data, from albino guinea pig, recorded with a dense 30-electrode array (30 $\mu m$ spacing, Multi Channel Systems MCS GmbH, Reutlingen, Germany).   After anesthesia with ketamine/xylazine (100/20 mg/kg) and pentobarbital (100 mg/kg), the eyeball was enucleated and the animal was killed by pentobarbital overdose in keeping with the AVMA guidelines on euthanasia. The eyeball was hemisected and the retina was allowed to dark adapt.   A small piece was cut out, separated from the pigment epithelium,  mounted (ganglion cells up) onto a piece of filter paper, and placed ganglion cells down onto the MEA.  A $15\times15$ flickering checkerboard consisting of binary noise, updated at 30 Hz, was projected onto the tissue.  We alternated between uncorrelated and exponentially correlated (space constant 50 $\mu m$; time constant 33 ms) stimuli.

Our procedure for identifying spikes in the recorded data had four steps: ({\bf 1}) Preprocessing (\sref{pp}), where spatial locality was exploited to segment the data, ({\bf 2})  Clustering and template building (\sref{ci}), where a subset of the data was clustered to separate the responses of likely neural units, template waveforms for each neural unit were built, and their variability characterized, ({\bf 3}) Spike fitting (\sref{sf}), where every firing event was separated into a superposition of responses from different neural units,  and ({\bf 4}) Validation of templates (\sref{ecr}), where each template and the spikes identified with it were tested for reliability.  

\subsection{STEP 1:  Data preparation and segmentation \label{s:pp}}

The first step in our procedure was to prepare the data for clustering of events from different neural units, by separating firing events from noise, and segmenting  spatio-temporally distinct regions of spiking activity on the electrode array.

Data from the array were sampled at 10KHz, high-pass filtered below 200 Hz with a finite impulse response filter to remove low frequency baseline fluctuations, and then packaged into  $3.2\,\msunit$ clips: (a)~``noise clips'' in which the potential never fell below \Nnoisedef, and (b)~``spike events'' surrounding moments at which the potential crossed \Nspikedef.   Clips with potentials between \Nnoisedef~ and \Nspikedef~ were neither used to characterize noise (since they might contain small spikes) nor used to identify spikes (since they were very noisy). The threshold for spikes was set to $\sim$4 times the standard deviation of the potential in the noise clips.  Each spike event thus consisted of $N=3.2\,\msunit\times10\,\kHzunit\times5\times6=960$  numbers, the potentials on a $32\times5\times6$ grid of space-time pixels (``stixels'').  Spike events sometimes overlapped each other, for example if a burst of spikes lasted longer than $3.2\,\msunit$.    Cluster identification and spike template building (\sref{ci}) used four 30-second segments sampled from different times, but subsequent spike fitting and sorting (\sref{sf}) used all the data.

Electrodes can share signals because  of instrumental cross-talk and because the activity of neurons spreads passively to nearby electrodes. Both effects can be captured by a linear filter that spatially blurs signals, and also applies to noise. Thus, we measured the spatial covariance of noise clips --- it was spatially isotropic and had a roughly exponential falloff, with a correlation length of $\sim 30\,\umunit$.  We applied the square root of the inverse of this covariance matrix to all data, and used the resulting ``spatially whitened'' data for all analysis. In some datasets this transformation sharpened the individual spikes spatially, improving our ability to distinguish them in the clustering stage. (In other datasets the transformation had little effect.) Our data also exhibited temporal correlations, but these have a different physical origin from the essentially instantaneous passive spatial spread.  We found that {\it temporal} whitening prior to clustering \cite{Pouzat:2002p1281} worsened our signal/noise ratio and impeded cluster determination.   Thus we incorporated temporal correlations later, during the spike fitting.

\paragraph{Segmentation: }Each spike event is a superposition of spikes from an unknown number of distinct neural units with stereotyped waveforms that we sought to identify.  We first spatially segmented the data to isolate waveforms from individual units and their immediate neighbors.   To this end, we identified all stixels at which the potential was more negative than the threshold of \Nspikedef{} and divided this set into connected components (two stixels were considered connected if they were nearest neighbors in either time or space). Within each connected patch we identified the absolute peak electrode and time, then extracted a $3.2\,\msunit$ region centered temporally on the peak time and cropped spatially to a neighborhood of nine channels surrounding (and including) the leader electrode. Thus each spike event was segmented into one or more cropped events; each of which was then classified according to its leader electrode. \footnote[1]{A similar segmentation method has recently been applied to the spike identification problem by J. Schulman (unpublished); see \texttt{http://caton.googlecode.com}.}  In  subsequent clustering,  only those events having the same leader electrode were directly compared to each other \cite{Litke:2004p1134}.   

Some cropped events might be composites of two spike types corresponding to neighboring, but distinct, neural units. However, this step at least decomposes composite events whose components are well separated in space or in time, and hence reduces the combinatorial burden inherent in large arrays; later steps handle composites missed at this stage. The method also ensures that, if spikes from two well-separated units frequently co-occur, the two units will nevertheless be correctly handled as separate.

\subsection{STEP 2: Cluster identification and template building\label{s:ci}}
The second step in our procedure was to cluster spiking events in a subset of the data (four 30-second segments) into groups that had similar waveforms and thus probably came from the same neural unit.   For each cluster, we produced a template waveform describing the typical spike, and determined the distribution of amplitude rescalings that best matched spikes to this template.

\paragraph{Cluster identification: }
In order to group events into clusters based on the similarity of their waveforms, some previous approaches have sought  a low-dimensional set of discriminable waveform ``features,'' and have assumed that variability between events in the same cluster arises only from additive noise.   In practice, systematic variation in the shape of spikes from single units is often observed that is not due to additive noise.   Furthermore, identifying the correct set of salient waveform features that discriminate between units is challenging (\cite{Quiroga:2007}; see \fref{3plot}B).  Thus, seeking a technique that did not require feature extraction,  we adapted the \optics{} algorithm \cite{Aanke99b}.  Briefly, \optics{} computes distances between all pairs of waveforms, then orders the waveforms such that similar ones are placed close together in a single linear sequence.   \optics{} makes no assumption that clusters have a Gaussian distribution in feature space, nor does it set any threshold density in that space to trigger cluster identification.  The linear ordering  allows for easy visualization and cutting of clusters.

We applied this algorithm to cropped and segmented spike events which were upsampled by a factor of 5 (using \ml's cubic spline interpolation) and then temporally aligned to place the absolute peak of the waveform at a common position before downsampling again.   The interpolation was necessary to compensate for apparent variations in spike waveforms due to discrete sampling  \cite{Lewicki:1994p1177}.  To reduce the fuzziness of the clusters, we masked spike events by setting voltage samples to zero if they were less negative than $-15 \mu$V.  As a distance metric between ${\bf V}$ and ${\bf V}'$, the masked potentials of spike events, we chose
$
d({\bf V}, {\bf V}') = \left( \sum_{i=1}^N [ (V_i - V'_i)^2 / (k \sqrt{|V_i| + |V'_i|}) ] \right)^{1/2}\, 
$
where $i$ indexes the potentials at each channel and timepoint, and $k$ is the total number of nonzero potentials after masking of either ${\bf V}$ or ${\bf V}'$.  Division by $k$ normalized for the effective dimensionality (given by the number of dimensions containing nonzero entries).  We observed that higher voltage traces tended to have a higher variance; the factor $\sqrt{|V_i| + |V'_i|}$ partially compensated for this, leading to more homogeneous clusters.

We  constructed a graphical user interface (\textsc{gui}) that allowed a human operator to visualize each spike event in the \optics{} sequence as a vertical column of pixels color-coded by voltage (see \fref{3plot}D). Transitions between distinct spike types were usually obvious to the operator, who could quickly find and select bands corresponding to each spike type. (For the data in this paper, the operator found over 100 such clusters in about 30  minutes.) The software then wrote the corresponding cropped events to a set of data files. The ease of separation likely occurred because clusters could already be fairly well delineated with just the ``geographical'' features in \fref{3plot}A.

Up to this point, the events being clustered were still segregated into batches according to their leader channel  $x_0,\,y_0$. Thus it was possible for a single unit to be multiply identified: If it stimulated two neighboring electrodes nearly equally, the unit could generate events in both of the corresponding batches. We tested for duplicates by manually examining pairs of clusters whose medians had a large cross-correlation and merged the clusters if necessary.  There was also a possibility that the initial clustering would assign multiple units to one cluster. In these cases, visual examination of the superposed waveforms of the cluster often showed it to be a composite of multiple units.   This was resolved by doing a principal components analysis on the waveforms in the cluster: if the cluster was composite, at least one of the first few principal component weights had a multimodal histogram.  The cluster was split by thresholding at the valleys of the histogram; we then tested whether any of the split components ought to be merged with an existing cluster. We developed a graphical user interface to assist the operator in performing these merging and splitting steps.

Generally it was clear to the human operator when a band in the \textsc{gui} output was clean enough and wide enough (contains enough events) to generate a good cluster; thus there was no need to specify \emph{a priori} the desired number of  clusters, an advantage over many automated clustering procedures. Marginally significant clusters were either eliminated during template building (see below), or else generated fits that were themselves discarded during spike fitting  (\sref{sf}) and evaluation of  template reliability (\sref{ecr}).  Any significant clusters \emph{missed} at this stage, for example because of the small fraction of the data used in this step, were found and reincorporated later during spike fitting (\sref{sf}).

\paragraph{Template building: }
Next we created  a consensus waveform (``template'') summarizing each cluster of cropped, upsampled events, and characterized meaningful deviations from that consensus.  We created a draft template by finding the pointwise median over all events in a cluster, then aligned each event to the draft template by maximizing their cross-correlation over time shifts, which we found to be more accurate than aligning to each event's peak time.
Finally, we found the pointwise median (to suppress the effects of outliers) of the aligned events; this waveform was our template (\fref{template}B).

\begin{figure}[tb!]
\begin{minipage}[b]{\textwidth}
\ifx\includefigs\yesflag
\begin{center}
\includegraphics[width=6.83in]{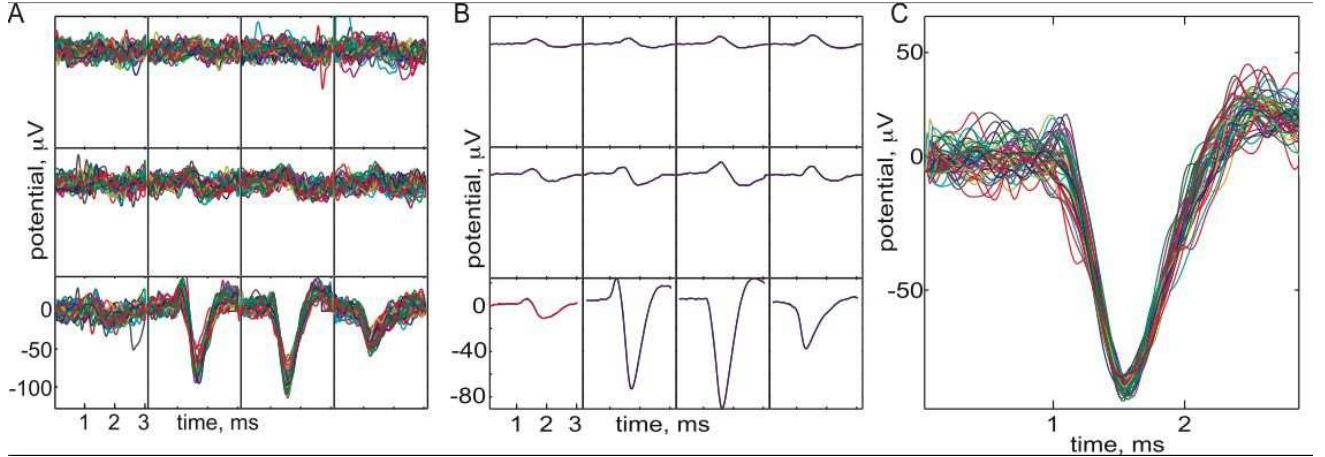}
\end{center}\fi
\end{minipage}
\hfil\renewcommand{\baselinestretch}{.95}\normalsize \smallskip \caption{\small
\capitem{A}\figstatus\draftfig{} Detail of 40 of the aligned events used to compute a template, upsampled and shifted into alignment as described in the text. 
Some outlier traces reflect events in which this unit fired together with some other unit; the unwanted peaks occur at random times relative to the one of interest, and thus do not affect the template. 
\capitem{B}{}\textsl{Blue,}~detail of a template waveform, showing the potential on 12 neighboring electrodes. Time in $\msunit$ runs horizontally; the vertical axis is potential in $\uVunit$.
\textsl{Red,}~for comparison, the pointwise mean of the 430 waveforms used to find this template. 
\capitem{C}\figstatus\draftfig{}Detail of (A), showing only the leader channel. In addition, each trace has been rescaled by a constant to emphasize their similarity apart from variation in overall amplitude. \pnlabel{f:template}}
\par
\end{figure}%

Our code displays 40 events in each cluster together, so that a human operator can spot any mixed (decomposable) cluster inadvertently missed by earlier stages of the analysis (\fref{template}A). Generally such clusters can safely be discarded, because each ``parent'' neural unit has also generated its own ``pure'' cluster; if not, the operator can either revisit the \optics{} code specifically to find the missed events, or else wait for them to show up during spike fitting (\sref{sf}).

A key step was to realize that, in our data, the most significant sources of variation of individual spikes from the template were (a)~additive noise, and (b)~overall multiplicative rescaling of the spike's amplitude (\fref{template}C). To quantify (b), we found the  rescaling factor that optimized the overlap of each spike with its template, then stored the mean and variance of those factors in a lookup table for later use as a prior probability for amplitude variation.  We also logged the number of events associated to each template, converted to an approximate firing rate, and saved those rates, again for later use as a prior.

\subsection{STEP 3: Spike fitting\label{s:sf}}

The third step in our procedure was to fit the spike templates constructed in Step 2 to each firing event in the data in order to determine which neural units were responsible for the activity.   To this end, we constructed a simple generative model of firing events, and  included prior probabilities for firing rate and for amplitude variation of each template.  The fitting procedure then iteratively identified and subtracted the most likely templates in each firing event.

The cluster templates were produced using an upsampled $50\,\kHzunit$ sample rate, but for fitting to data we  downsampled back to actual $10\,\kHzunit$, in each of 5 ``reading frames'';  that is, we created five versions of each template corresponding to subsample shifts.   Let $F_{\mh;x,y}(t)$ be the potential of template $\mh$, on the electrode with address $x,y$, at time $t$, with time measured in units of the sampling time $\delta = 0.1\,\msunit$, and the template peak at the central point $t=16$ within the 3.2 ms template frame.  We use the vector notation  $\bF_{\mu_i t_i}$ for the template $\mu_i$ shifted to time $t_i$, i.e. its $x,y,t$ component is $F_{\mh_i,x,y}(t-t_i)$.

\paragraph{Generative model: }
 The goal of spike fitting is to identify, for each spike event, all the units $\{\mu_i\}$ which contribute to the event and their firing times $\{t_i\}$ irrespective of their amplitudes $\{ A_i \}$.  Thus we assumed a probabilistic generative model of the data \cite{ atiya1992recognition, Lewicki:1994p1177,Tsaha99a,Pouzat:2002p1281} and computed the posterior probability of $\{(\mu_i,t_i)\}$ given the observed data.   We assumed that a spike event $\bV$ could be explained by a linear combination of templates $\bF_{\mu_i t_i}$ with variable amplitudes $A_{i}$ and correlated, zero-mean Gaussian noise $\delta\mathbf{V}$:
\begin{equation}
\bV = \sum_{i} A_{i} \bF_{\mu_i t_i} + \delta\mathbf{V}.
\label{e:genmod1}
\end{equation}
Given this model, to obtain the posterior probability that a firing event consists of a particular set of templates, we need to specify the prior probability of $\mu$, $t$, and $A$. We chose a Gaussian prior for the amplitude $A$, a Poisson prior for $\mu$, and a uniform prior for $t$.  

Our model assumes that spike waveforms from a given neural unit are stereotyped, apart from their amplitude.  We did observe considerable variation in spike amplitude (\fref{3plot}A), in part due to bursting  \cite{Fee:1996p1393,Harris:2001p2555}, and thus included it in the model as a distribution of amplitude rescaling factors.   Allowing for the possibility of slight variations in spike width  (Supplementary Information) also slightly improved our results.  But there was little additional variability  to be modelled (\fref{template}C).   Our model also assumes that signals from different units combine linearly, as does the noise. This is reasonable, because the biophysics of extracellular recording is governed by the equations of electrodynamics, Ohm's law, and other linear relations.   A third assumption is that noise and the variability of spike amplitude are well described by Gaussian distributions.  Assuming Gaussianity (well-confirmed in some settings  \cite{Pouzat:2002p1281}, but not others \cite{Shoham:2003p1275}) allows for a fast, partially analytic approach to fitting.   We validate this assumption quantitatively below.    

Our generative model has a Poisson prior probability for firing by each neural unit, i.e. a prior that is as simple as possible while being consistent with the mean firing rate.   The prior probability could be made somewhat more accurate by including refractory periods, the likelihood of bursting, and correlations between neural units.   But this would significantly increase the complexity of the model, and inferring the prior would require much more data \cite{Schneidman:2006p1088}.   

 Finally, we assumed  that  all statistical distributions that enter into the model are stationary and independent of the stimulus.   While  our retinal preparation does not suffer electrode drift (as might implanted electrodes), there are occasionally shifts in spike amplitudes and firing rates over the course of a lengthy experiment.   Although in principle our fixed priors  could lead to biased estimates,  these biases are small when spike identification is robust, i.e. when the likelihood function dominates the prior probability in the posterior probability of a neural unit~\cite{Ventura:2009p1257}.


\paragraph{Noise characterization: } In the context of our generative model, in order to assess the probability that the residual after subtracting a putative spike is indeed noise, we first need to measure the distribution of noise. 
 After applying the spatial whitening filter (\sref{pp}), our noise clips are decorrelated in space, but  not in time (\fref{testnoise}A). Assuming that the noise has a correlated Gaussian distribution, we need the inverse of the noise covariance matrix, $\Ncov\inv$. One approach to finding $\Ncov\inv$ is literally to invert the empirical covariance matrix $\Ncov$ of a large set of noise clips. Besides being intractable for larger arrays, this approach has the disadvantage that a numerically stable evaluation requires a very large noise sample. 

For these reasons we instead took a parametric approach. After evaluating the covariance $\Ncov(x,y,t; x',y',t')$ we noted that it was approximately diagonal and translation-invariant in space (i.e. proportional to $\delta_{x,x'} \delta_{y,y'}$ and independent of $x$ and $y$).   It was also approximately stationary, i.e.\ invariant under time shifts $t\to t+\Delta t$,  $t'\to t'+\Delta t$, and thus only depended on $t - t'$.   Finally, we observed that the time dependence of $\Ncov$ was roughly an exponential, $\approx\eta\ex{-|t-t'|/\tau}$ (\fref{testnoise}A).  This gave:
\begin{equation}
\Ncov\inv(x,y,t;x',y',t')=\eta\inv\times\delta_{x,x'}\delta_{y,y'}\left\{
\begin{array}{r@{\ \mathrm{,\ }\ }l}
(1+\Nfall^2)/(1-\Nfall^2)&\mathrm{if\ }t=t'\\
-\Nfall/(1-\Nfall^2)&\mathrm{if\ }t=t'\pm\delta t\\
0&\mathrm{otherwise}.
\end{array}
\right.
\label{e:Chati}\end{equation}
Here $\delta t=0.1\,\msunit$ is the sample time, and $\delta_{x,x'}$ is the Kronecker symbol. 
$\eta$ and $\Nfall=\ex{-\delta t/\tau}$ are obtained from the noise covariance. The dataset used in Results yields noise strength $\eta\approx\Neta$ and $\Nfall\approx\Nx$.  

By construction, our noise model  reproduces the  2-point correlations in the noise clips. However, real noise may not be Gaussian distributed.  One check on this is to construct the transformed quantities ${\bf U}=\Ncov^{-1/2}\bV$  empirically, find their full distribution as $\bV$ ranges over noise clips (the ``one-point marginal'' distribution), and compare to a normal distribution. \fref{testnoise}B shows this comparison, lumping together every element of $\bf U$. The empirical noise deviates from a Gaussian only in the far tails that contain very little weight.

\begin{figure}[tb!]
\begin{minipage}[b]{\textwidth}
\ifx\includefigs\yesflag
\begin{center}
\hspace{0in}
\includegraphics[width=6.83in]{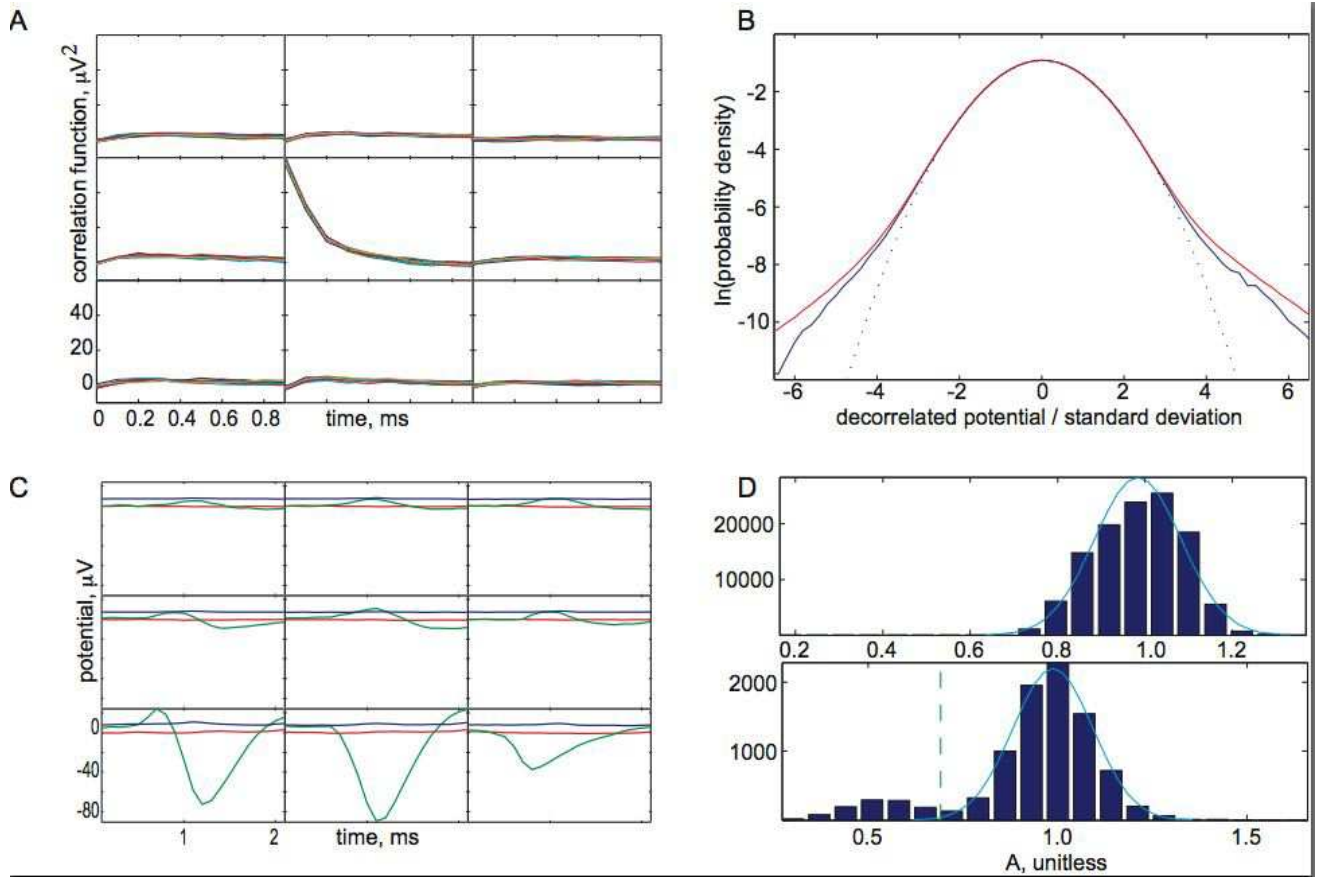}
\end{center}\fi
\end{minipage}
\hfil\renewcommand{\baselinestretch}{.95}\normalsize \smallskip \caption{\small
After fitting spikes, only noise remains.
\capitem{A} Noise covariance after  spatial whitening. Subpanels:   spacetime covariance  $\Ncov(x,y,t; x_*,y_*,t+\Delta t)$ 
between the central channel  and its neighbors as a function of $\Delta t$, for various fixed $t$ (colored curves).   Central panel (dotted line): the function $(\NfitCorMag)\exp(-\Delta t/(\NfitCorLen))$.  (The various $t$ lines and the dotted line are too similar to discriminate visually.) Horizontal axes: $\Delta t$ in \msunit; Vertical axes: $\Ncov$ in $\uVunit^2$.  
 \capitem{B}  \capcolor{Blue curve,}~Semilog plot of the 
one point marginal probability density function of decorrelated noise samples. 
\capcolor{Red curve,}~same quantity, evaluated on residuals after spikes have been removed from spike events.
\capcolor{Dotted curve,}~The Gaussian chosen to represent this distribution.
\capitem{C}
\capcolor{Green,}~detail of the same template waveform shown in \fref{template}.
\capcolor{Red,}~pointwise mean of the residuals after the fit spike is subtracted from 4906 one-spike events  of this type is nearly flat.  This validates our assumption that spikes vary only in overall amplitude, and that  noise is independent of spiking.
\capcolor{Blue,}~pointwise standard deviation of the residuals, again evidence that only noise remains after fitting and subtracting spikes.
\capitem{D, top} Histogram of fit values of the scale factor $A$ for a template with peak amplitude $-168\,\uVunit$ (well above noise) obtained without a prior on $A$, superposed with a Gaussian of the same mean and variance.  
\capitem{D, bottom} Similar histogram for a low amplitude template.  A secondary bump appears, due to noise-fits, but is well separated from the main peak; a cutoff is shown as a dashed green line. The superposed Gaussian has mean and variance computed from the part of the empirical distribution lying above the cutoff. 
 \pnlabel{f:testnoise}}\par
\end{figure}%

\paragraph{Fitting algorithm for single spikes: }  Given the above characterization of the noise distribution, and our Gaussian prior for spike amplitude variation, the generative model \eref{genmod1} defined the posterior probability $P(\{\mh_i,t_i,A_i\}|\mathbf{V})$ for templates $\{\mu_i\}$ to be present at times $\{t_i\}$ with amplitude scale factors $\{A_i\}$, given the recorded potentials $\mathbf{V}$.  We ideally would have marginalized  $P(\{\mh_i,t_i,A_i\}|\mathbf{V})$ over the nuisance parameters $\{A_i\}$  and then maximized with respect to $\{(\mu_i,t_i)\}$ to identify the most probable set of units and spike times.  In practice, this maximization is computationally expensive to perform on many templates simultaneously.  Instead, we used a greedy approximation which fit one template at a time.  

We first assumed that the event contained exactly one spike and identified the spike's type $\mh_1$ and time of occurrence $t_1$.   Bayes' formula gives for the posterior probability:
\begin{equation}
P(\mh_1,t_1,A_1|\mathbf{V})\dd t_1\dd A_1 \propto P(\mathbf{V}|\mh_1,t_1,A_1)P(\mu_1,t_1,A_1)\dd t_1\dd A_1 \, ,
\end{equation}
up to a constant independent of $\mu_1$, $t_1$, and $A_1$.   Here $P(\mu_1,t_1,A_1)$ is the prior probability of the template $\mu_1$ appearing at time $t_1$ with an amplitude $A_1$:
\begin{equation}
	P(\mu_1,t_1,A_1)\dd t_1\dd A_1 = r_{\mu_1}\dd t_1(2\pi{\sigma_{\mu_1}}^2)^{-1/2}
	\exp\left(-(A_1-\gamma_{\mu_1})^2/2\sigma_{\mu_1}^{2}\right)\dd A_1 \, ,
\label{e:prior}\end{equation}
where $\gamma_{\mu_1}$ is the mean and $\sigma_{\mu_1}^2$ the variance of the scale factor for cluster $\mu_1$; $r_{\mu_1}$ is the estimated overall rate of firing for this cluster. 
The generative model gave the probability of the observed potential $\mathbf{V}$ given $\mu_1, t_1, A_1$ (the likelihood) as
$
	P(\mathbf{V}|\mh_1,t_1,A_1)=P_{\rm noise}(\mathbf{V} - A_1\bF_{\mh_1,t_1}) \, ,
$
where $P_{\rm noise}$ is a Gaussian distribution with zero mean and covariance $\Ncov$ (\eref{Chati}). Combining the likelihood and prior, then integrating out $A_1$, gave the formula we ultimately used in our fitting algorithm:
\begin{equation}
P(\mu_{1},t_1 \, | \, \mathbf{V}) \propto \frac{r_{\mu_1}}{\sqrt{ 1 + \sigma_{\mu_1}^{2} \bF_{\mu_1,t_1}^{\mathrm{t}} \Ncov\inv \bF_{\mu_1,t_1}}} \exp \left(\frac{1}{2} \frac{(\gamma_{\mu_1} + \sigma_{\mu_1}^{2} \bV^{\mathrm{t}} \Ncov\inv \bF_{\mu_1,t_1})^{2}}{ 1 + \sigma_{\mu_1}^{2} \bF_{\mu_1,t_1}^{\mathrm{t}} \Ncov\inv \bF_{\mu_1,t_1}} - \frac{\gamma_{\mu_1}^{2}}{2 \sigma_{\mu_1}^{2}}\right).
\label{e:margpost}\end{equation}
The (unwritten) constant of proportionality in (\ref{e:margpost}) is the probability that \textit{no} templates are present in the event; this quantity cancelled in a subsequent step.  Finally, we maximized (\ref{e:margpost}) over $\mh_1$ and $t_1$ to identify the template and its firing time. We improved scalability by a slight approximation. Starting from a spike event, we first identified the time and electrode address of its absolute peak and restricted the matrix products in expression (\ref{e:margpost}) to only sum over a spatiotemporal neighborhood surrounding this peak.  The size of the neighborhood was chosen to match the typical spatial extent and temporal duration of the templates.

\paragraph{Multiple spikes: }
In principle, we could have extended the single template procedure described above to compare the probabilities of all possible combinations of two or more spikes. Such an exhaustive approach, however, would quickly have become impractical. We instead noted that, even if an event contains multiple spikes, the single-spike fit described above still identified that template whose removal would lead to the largest increase in the probability that the remaining waveform is noise. Thus  we  adopted an iterative (matching-pursuit or ``greedy'') approach: starting with a spike event, we found the absolute peak, fit it,  subtracted the fit, and then repeated the process \cite{Segev:2004p1133}.  The single-spike procedure found the most probable spike type $\mh_*$; we then needed the scale factor $A$ that would allow the fit spike to be subtracted as fully as possible. We thus held $\mh$ fixed to $\mu_*$ and minimized the ordinary norm $\|\bV-A\bF_{\mh_*,t_1}\|^2$ over $A$ and $t_1$. The scaled and shifted template obtained in this way was subtracted before repeating the fitting procedure.

To determine when to stop fitting spikes, we adopted a likelihood ratio test. At each step of the fitting loop, we marginalized \eref{margpost} over $t_1$, obtaining the probability that an additional spike of type $\mu_1$ is present. We then divided by a similar expression for the probability that \emph{no} additional spike was present.  In this ratio, the proportionality constant from \eref{margpost} cancels.  We can then say that fitting an additional spike is justified if the ratio exceeds unity for some $\mh_*$. The fitting loop terminates when the significance test fails.  \fref{expt}C,D shows an example of the successful decomposition of a multiple-spike event using our method.

%

\paragraph{Second pass: } 
The spike fitting algorithm might exit prematurely if a spike is present that does not appear in the list of templates initially extracted from the small subset of data.  In this case, the fit will terminate, even though other identifiable spikes of lower amplitude may remain. To check this, if the residual exceeds $\NinterestingVThreshsym=$\NinterestingVThresh~ after termination, the code declares an ``incomplete fit'' and writes the residual to a file; the  small set of resulting waveforms were then  reintroduced into our clustering code and used for a second round of fitting. In this way we can be assured of finding even rare spike types, without having to perform clustering on the complete dataset. Using this method, only $\NIFpercent\%$ of fits in the second pass were classified as incomplete.
It can also happen that the small data sample used for clustering gives a poor estimate of some firing rates and amplitude distributions that enter our priors for spike fitting.  Thus, before the second pass of fitting the priors are updated based on the outcome of the first pass.

\subsection{STEP 4: Evaluation of template reliability\label{s:ecr}}
After spike identification, we performed a final evaluation to test whether templates and their sorted spike trains were trustworthy.  The primary criteria were: (1) residuals after spike removal should resemble noise, (2) the histogram of amplitude scale factors should be unimodal,  (3) the inter-spike interval (ISI) distributions should display ``refractory holes''; and the cross-correlation functions should not. Additional criteria are described in the Supplement. Reliable templates were taken to be those that passed all these tests. Most unreliable templates failed multiple tests.

\paragraph{Residuals: } For single-spike events, the residual signal after subtracting the fit should resemble pure noise; in particular it should be stationary in time and translation-invariant in space.  \fref{testnoise}C shows that these expectations were met, validating our assumptions.  For example, if the unit in  \fref{testnoise}C had significant variations other than amplitude rescaling, or if there had been an amplitude-dependent noise process, then we would expect significant non-stationarity in the residual curves \cite{Lewicki:1994p1177}. To test that, after termination, the residual of a spike event consists only of noise, we computed the one-point marginal distribution of waveforms after all known spikes had been removed.  \fref{testnoise}B shows that this  distribution closely resembled that of noise clips, indicating that our code indeed found the significant spikes. Of particular note, the standard deviation of the residuals matched closely that of the noise: For each template, we found the standard deviation of the residuals of events to which only that template was fit. This value ranged from $6.92$  $\mu$V to $8.63$ $\mu$V, while the noise standard deviation was $7.57$ $\mu$V.

\paragraph{Amplitude\label{s:etra}: } For large amplitude templates, the distribution of amplitude scale factors ($A$) obtained during spike fitting was typically close to Gaussian (\fref{testnoise}D).     On the other hand, for low amplitude templates, accidental noise fits can sometimes lead to a histogram of $A$ values with a secondary, low-amplitude peak well separated from the expected peak near $A\approx1$ (\fref{testnoise}D). Examining the $A$ histograms allowed us to quickly set an individual threshold for each reliable template. Fit spikes with $A$ value below this threshold were discarded. If two peaks were discernible but overlapped significantly, the entire cluster was deemed unreliable and its spikes were not used in further analysis. In addition, our trigger rejected any spike event that did not cross \Nspikedef; thus any cluster whose $A$ histogram extended closer to zero than this was probably missing some true spikes, and was not used.

\paragraph{ISI distribution and cross-correlation\label{s:etrcc}:} Interspike interval distributions for single units are expected to have a refractory hole; our analysis of these distributions was described in \ref{s:tests}. Two distinct neural units need not respect any mutual refractory period. Their spike-time cross-correlation function is therefore \emph{not} expected to display any hole.  We looked for such unexpected behavior and, when found, reexamined the corresponding templates. If the templates appeared to be duplicates, we merged the corresponding spike trains \cite{Litke:2004p1134,Segev:2004p1133}. Another diagnostic for duplicate templates is a coincident receptive field.

\section{Discussion\label{s:disc}}
A review of early work on spike identification, prior to the widespread use of MEAs, can be found in \cite{Lewicki:1998p1226}.  Like some earlier work, our method separates spike identification into distinct steps of clustering and fitting. The clustering step  uses all the waveform features, and makes no assumption about the cluster structure (e.g., that it is a mixture of Gaussian distributions). The fitting step acknowledges that each neural unit's signals are subject to intrinsic, multiplicative variation as well as additive noise, and uses a Bayesian approach to infer the identity of the most likely firing unit.  

Our approach is intentionally not fully automatic, since human proof-reading of the results of automated clustering is generally essential.  However, we have been careful to use human judgement only where it is indispensable.  Further, both the human and machine steps are organized so as to scale well with array area (or number of units monitored).   For example, cluster cutting was greatly simplified by representing spikes in an ordered one-dimensional array.   This feature, along with systematic exploitation of the spatio-temporal locality of spikes (\fref{3plot}A), and the use of a simple but powerful generative model, make our method scalable to large arrays. Furthermore, we observed that our more ambiguous templates tended to be located on the boundary of the array due to recording of units located some distance from the electrodes. These ``boundary effects'' should become less important for larger arrays; we thus anticipate that the methods described in this paper will yield more accurate spike sorting for large, dense arrays.

Our method can be extended in many ways. For example, it would be straightforward to update the priors continually as fitting proceeds, allowing non-stationarity and stimulus dependence to be handled more gracefully. In some applications it may be preferable to report spike identification probabilistically, rather than just listing the most-likely spike events; our formulas already provide this information.   The method can also be extended to non-planar arrangements of electrodes and neural tissues, for, e.g., cortical applications.   Finally, the generative model in the present paper does not take into account correlations within and between spike trains, or the receptive field structure and stimulus dependence of responses.    While our algorithm is already very accurate, performance could be improved on complex overlapping spike events via a bootstrapping procedure.  We could first use the simple independent, Poisson generative model of this article to produce an accurate preliminary assignment of spikes to units.  From this assignment we could construct a more detailed model of correlated activity with pairwise interactions (e.g.  \cite{Schneidman:2006p1088} or the stimulus-dependent models \cite{Pillow:2008p654,isingPNAS}).   This more complex generative model could then be used to further refine spike assignments for applications requiring a very high degree of accuracy.

\section*{Acknowledgments}
We thank  Michael Berry, Michael Freed, Mike Jarvis, Olivier Marre, Liam Paninski and Joerg Sander for helpful comments. This work was supported by  NSF grants IBN-0344678,  EF-0928048, NIH grant RO1 EY08124, NIH training grant T32-07035, and NIH training grant 5T90DA022763-04.

\bibliography{sortFoot,spikeSort,sortExtras}

\end{document}